\documentclass[prb,amssymb,amsmath,twocolumn,showpacs]{revtex4}
\usepackage{amsfonts}
\usepackage{amsmath}
\usepackage{amssymb}
\usepackage[english]{babel}
\usepackage{fontenc}
\usepackage{graphicx}
\usepackage{color}
\usepackage{bbm}
\usepackage{natbib} % bibliography style

\newcommand{\sgn}{\mathrm{sgn}}

\makeatletter
\newcommand{\ulinks}[3]{%
  \@mathmeasure\z@\displaystyle{#3}%
  \global\setbox\@ne\vbox to\ht\z@{}\dp\@ne\dp\z@
  \setbox\tw@\box\@ne
  \@mathmeasure4\displaystyle{\copy\tw@#1}%
  \@mathmeasure6\displaystyle{#3#2}%
  \dimen@-\wd6 \advance\dimen@\wd4 \advance\dimen@\wd\z@
  \hbox to\dimen@{}\mathop{\kern-\dimen@\box4\box6}%
}
\makeatother
\newcommand{\bp}{\begin{pmatrix}}
\newcommand{\ep}{\end{pmatrix}}

\newcommand{\measure}{\mathrm{d}}

\begin{document}
\title{Absence of split pairs in the cross-correlations of a highly
  transparent normal metal-superconductor-normal metal electron beam splitter}

\author{Martina
  Fl\"oser}

\affiliation{Univ. Grenoble Alpes, Inst NEEL, F-38042 Grenoble, France}
\affiliation{CNRS, Inst NEEL, F-38042 Grenoble, France}

\author{Denis Feinberg}

\affiliation{Univ. Grenoble Alpes, Inst NEEL, F-38042 Grenoble, France}
\affiliation{CNRS, Inst NEEL, F-38042 Grenoble, France}

\author{R\'egis M\'elin}
\email{regis.melin@grenoble.cnrs.fr}

\affiliation{Univ. Grenoble Alpes, Inst NEEL, F-38042 Grenoble, France}
\affiliation{CNRS, Inst NEEL, F-38042 Grenoble, France}

\begin{abstract} 
The nonlocal conductance and the current cross-correlations are investigated
within scattering theory for three-terminal normal metal-superconductor-normal
metal (NSN) hybrid structures. The positive cross-correlations at high
transparency found by M\'elin, Benjamin and Martin [Phys. Rev. B {\bf 77},
  094512 (2008)] are not due to crossed Andreev reflection. On the other hand,
local processes can be enhanced by reflectionless tunneling but this mechanism
has little influence on nonlocal processes and on current
cross-correlations. Therefore Cooper pair splitting cannot be enhanced by
reflectionless tunneling.  Overall, this shows that NSN structures with highly
transparent or effectively highly transparent interfaces are not suited
to experimentally producing entangled split pairs of electrons.
\end{abstract}
\maketitle

\section{Introduction}
Transport in normal metal-superconductor-normal metal (N$_a$SN$_b$)
three-terminal hybrid nanostructures has received a special attention, because
those structures allow in principle to produce split pairs of spin-entangled
electrons from a superconductor, acting as a Cooper pair beam splitter\cite{recher,lesovikmartin}. This is possible
when the size of the region separating the N$_a$S and the N$_b$S interface
becomes comparable to the superconducting coherence length, allowing coherent
processes involving two quasi-particles, each simultaneously crossing one of
the two interfaces
\cite{allsopp,byers,torres,deutscher,falci,melin1,melin2}. Much effort has
been devoted to the theoretical understanding and to the experimental
observation of such a Cooper pair splitting effect. In a transport experiment
where electrons are difficult to measure one by one --- contrarily to the
similar production of entangled photons ---, one relies on steady transport
measurement, e.\,g. the current-voltage characteristics (conductance) and the
cumulants of the current fluctuations (non-equilibrium current noise
\cite{antibunching,bignon} and its counting statistics
\cite{fazio,morten}). In practice, the conductance and the second-order
cumulant (the shot noise and the current cross-correlations between the two
current terminals N$_a$, N$_b$) are the quantities to be extracted from
experiments. Indeed, due to Fermi statistics, the ``partition`` noise
correlations at a three-terminal crossing of normal metal contacts are
negative~\cite{antibunching,AD1996}, manifesting the antibunching properties
of individual electrons. If instead one contact is made superconducting, the
cross-correlations may become positive, suggesting the splitting of Cooper
pairs~\cite{torres}.

Two elementary nonlocal processes occur at a double NSN interface: Crossed
Andreev Reflection (CAR) which alone leads to Cooper pair splitting into
separated electrons bearing opposite spins (for a spin singlet
superconductor), and Elastic Cotunneling (EC) which alone leads to
(spin-conserving) quasi-particle transmission between the normal contacts,
across the superconducting gap \cite{falci}. For tunnel contacts, at lowest
order in the barrier transparencies, those two processes are decoupled and
simply related to the conductance, leading to positive (resp. negative)
conductance and current cross-correlations for CAR (resp. EC)
processes. Indeed Bignon \textit{et\,al.}~\cite{bignon} showed that for tunnel
contacts the linear dependence of the current cross-correlation on the
voltages applied to the contacts N$_a$, N$_b$ allows to separately track the
amplitudes of CAR and EC. Due to the expected compensation of the opposite CAR
and EC conductance components at low transparencies
\cite{falci,brinkmangolubov} , ferromagnetic contacts are required to detect
CAR and EC from the conductance with tunnel contacts
\cite{deutscher,falci,sanchez}. Yet, such polarizations are not easily
achievable, moreover, if one is interested in producing spin-entangled
electrons in a nonlocal singlet state, one should of course not spin polarize
the contacts.

As regards experiments, the situation for (extended) tunnel barriers looks
more complicated than given by a simple tunnel model
\cite{DelftExpt,Levy-Yeyati}. 
%%%%%%%%%%%%%%%%%%%%%%%%%%%%%
%%%%%%%%% CHANGE 1  %%%%%%%%%
%%%%%%%%%%%%%%%%%%%%%%%%%%%%%
Zero-frequency noise measurements can be carried out in low impedance sample
(current noise), or at high impedance (voltage noise). The positive
current-current cross correlations discussed here at high transparency may be
measured in the setup of Ref.~\onlinecite{Lefloch} using three SQUIDs as
current amplifiers. It has been found theoretically that, at high transparency
\cite{melin2,melin3,duhotmelin,brataas,kalenkovzaikin,FFM2010}, the
%%%%%%%%%%%%%%%%%%%%%%%%%%%%%%%%%%%
%%%%%%%%% END OF CHANGE 1 %%%%%%%%%
%%%%%%%%%%%%%%%%%%%%%%%%%%%%%%%%%%%
nonlocal conductance is negative, which leaves the current cross-correlations
as the only possible probe of Cooper pair splitting processes, provided one
controls the voltages on both contacts. Contrarily to conductance measurements
with metals \cite{exptconductance} or quantum dots \cite{exptdots},
cross-correlations have led to few experimental results
\cite{exptcrossnoise,das2012}. At the theoretical level, the dependence of the
cross-correlations on the contact transparency is not yet fully understood.

In view of the current experiments on metallic structures, the main question
is therefore: Can the cross-correlations be positive, and if the answer is
yes, is this a signature of Cooper pair splitting ? Previous work on a NSN
structure~\cite{melinbenjamin} showed that the cross-correlations can indeed
be positive at large transparencies, although the nonlocal conductance is
negative.  The origin of this somewhat surprising result was not fully
elucidated. Further work~\cite{FFM2010} showed that the sign of the
cross-correlations indeed changes with the transparency of the interfaces,
being positive at low transparency, negative at intermediate transparency and
positive again at high transparency. While the positive sign at low
transparency is clearly ascribed to Cooper pair splitting, it was shown that
the positive sign at high transparency should not be interpreted in the same
way. Indeed, at high transparency, CAR processes do not dominate either in the
conductance or in the noise. Instead, the positive cross-correlations should
be ascribed to local Andreev reflection (AR) on one side, and the opposite
process on the other side, a process equivalent to exchanging a pair of
electrons between the two normal contacts. In
Ref. \onlinecite{ZaikinPRBnoise}, a quasiclassical analysis using a
perturbative expansion in the nonlocal Green's function connecting the two
interfaces led to positive noise correlations at high transparencies, and the
authors conclude that it is due to CAR. This interpretation looks surprising,
given the domination of EC in the conductance in the same regime, and the
total absence of CAR at high transparency in the transmission coefficient. We
insist that it is of importance for the community, before embarking into
experimental developments, to state clearly that positive cross-correlations
should not be interpreted in terms of CAR at high transparencies.

To show that contacts with high transparency, or with effectively high
transparency, are not suitable as a source of Cooper pair splitting, we also
investigate how the localizing effects of disorder influence the nonlocal
conductance and the cross-correlations in a NSN structure. At a single NS
interface, it was shown that disorder in the N region, or multiple scattering
at a clean NN$_l$S double interface, can strongly enhance Andreev reflection,
by a mechanism nicknamed ''reflectionless tunneling''
\cite{reflectionlesstunnexpt,reflectionlesstunntheo,MB1994}. With a disordered
$N_l$ region, it leads to a zero-bias anomaly below the Thouless energy, where
the mutual dephasing of electrons and Andreev-reflected holes is
negligible. In the case of a clean double NN$_l$S interface, maximum Andreev
transmission is obtained by balancing the transparencies T$_{nn}$ and T$_{ns}$
of the NN$_l$ and N$_l$S interfaces, such that the N$_l$ region acts as a
resonant cavity. Melsen and Beenakker \cite{MB1994} performed an average on
the modes inside N$_l$ in order to mimick a disordered region. We ask here the
question of whether a similar mechanism can enhance nonlocal Andreev
reflection e.g. boost CAR compared to AR and EC. A zero-bias anomaly was
obtained in Ref.~\onlinecite{ZaikinPRL}, within a quasiclassical analysis. Its
sign reveals an amplification of quasi-particle transmission. It is however
not mentionned in this reference whether one should also expect boosting of
the CAR channel, which is a requirement for experimental observation of Cooper
pair splitting. We demonstrate on the contrary that reflectionless tunneling
in the nonlocal conductance is not accompagnied by reflectionless tunneling in
the CAR channel.

The needed clarification, both for conductance and noise, comes from a model
which is exactly solvable and where all local and nonlocal amplitudes can be
clearly distinguished.  This is an advantage over the quasiclassical approach
used in Ref.~\onlinecite{ZaikinPRL}, which does not produce separate
expressions for the CAR and for the EC contributions to the conductance. Thus,
we use the scattering approach \cite{BTK1982} for a set-up
N$_a$N$_l$SN$_r$N$_b$ with a quadruple interface. The scattering theory is
performed in a one-dimensional geometry, varying the transparencies of the
barriers and the width of the superconductor. It is known to reproduce the
main qualitative features of realistic devices, and allows to account for any
barrier transparency and any distance $d$ between the interfaces. It does not
rely on any expansion in the nonlocal scattering matrix elements (or Green's
functions). This is especially important if noticing that close to the gap
edges, the relevant length scale for the penetration of evanescent
quasiparticles (thus for the Andreev reflection) diverges as $\xi(\omega) =
\xi_0 / \sqrt{1-\frac{\omega^2}{\Delta^2}}$. The scattering method only
assumes a sharp variation of the order parameter at the interface, which is
strictly valid for contact sizes smaller than $\xi_0$. It should be modified
to take into account self-consistency if $d \sim \xi$, or to describe
non-equilibrium effects.  One advantage of the scattering approach is the
precise bookkeeping of the scattering amplitudes associated to the
various (AR, CAR, EC) processes. This allows an unambiguous diagnosis of
Cooper pair splitting in either the conductance or the noise, as obtained by
simple expressions of these scattering amplitudes.

If averaging independently the modes in the left and the right regions N$_{l,r}$,
no boosting of the CAR process is obtained. Indeed, ``reflectionless
tunneling`` is a quantum coherent process which demands that the
Andreev-reflected hole retrace the path of the electron, by scattering on the
same impurities. On the contrary, with nonlocal Andreev reflection, the
electron and the transmitted hole sample different disorders and no coherence
is obtained. This result does not contradict Ref. \onlinecite{ZaikinPRL} which
states that the total crossed conductance is enhanced. Again, the
scattering technique allows to track the different contributions, and the
zero-bias anomaly is here due to the enhancement of the direct Andreev
reflection, not to CAR.

Section~\ref{sec:paperNSN:model} presents the model and
section~\ref{paperNNSNN:sec:comp} the scattering theory of the
N$_a$N$_l$SN$_r$N$_b$ system. Section~\ref{paperNNSNN:sec:AS} provides the
results obtained by averaging over channels in N$_l$ and N$_r$, in the spirit
of Ref.~\onlinecite{MB1994}. 

%%%%%%%%%%%%%%%%%%%%%%%%%%%%%%%%%%%%%%%%%%%%%%%%%%%%%%%%%%%%%%%%%%%%%%%%%%%%%%%%%%%%%%%%%%%%%%%%%
\section{The model\label{sec:paperNSN:model}}
\begin{figure}
\includegraphics[width=0.45\textwidth]{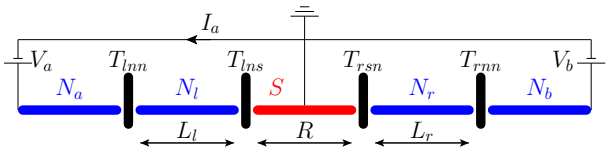}
% SchematicNSN.eps: 0x0 pixel, 300dpi, 0.00x0.00 cm, bb=-13 -24 212 64
 \caption{\label{paperNNSNN:fig:model} Schematic of the model.}
\end{figure}

We study a one-dimensional model of a symmetrical three-terminal normal
metal-superconductor-normal metal hybrid structure depicted in
Fig.~\ref{paperNNSNN:fig:model}. The central superconducting electrode is
grounded, the normal terminals are biased with voltages $V_a$ and $V_b$.  The
length $R$ of the superconducting electrode can be comparable to the
superconducting coherence length. The interfaces between the normal metal and
the superconducting electrodes are modeled by barriers with transparencies
$T_{lns}$ and $T_{rsn}$. In both normal metal electrodes there is an
additional barrier at a distance $L_l$ (respectivly $L_r$) from the normal
metal superconductor interface with transparency $T_{lnn}$ (respectivly
$T_{rnn}$).\\

The system can be described by a $4\times4$ scattering matrix
$s_{ij}^{\alpha\beta}$ where Latin indices run over the normal electrodes $a$
and $b$ and Greek indices over electrons $e$ and holes $h$. The scattering
theory assumes that the superconductor is a reservoir of Cooper pairs, so the
structure is implicitly a three-terminal one and the superconducting electrode
is taken as grounded. The transformation of quasi-particles into Cooper pairs
is taken into account by the correlation length $\xi(\omega)$ which sets the
scale of the damping of the electron and hole wavefunctions in the
superconductor.

The elements of the scattering matrix are evaluated from the BTK
approach\cite{BTK1982} (see Appendix~\ref{paperNSN:app:BTK}). Here we do not
use the Andreev approximation valid in the limit of zero energy where the
electron and hole wavevectors are set to the Fermi wavelength, but instead
keep the full expressions for the wavevectors.  The calculation of the average
current $I$ and current cross-correlations $S$ rely on the formulas derived by
Anantram and Datta in Ref.~\onlinecite{AD1996}:
%Formula for the operator, but we need only the expectation value.
%\begin{align}\label{paperNSN:eq:currentOp}
 %\hat{I}_i=e\sum_{k,l\in \{a,b\}}\sum_{\alpha,\gamma,\delta\in\{e,h\}} \sgn(\alpha)\int dE\phantom{0} A_{k\gamma,l\delta} %(i\alpha,E)
%\hat{a}^{\dagger}_{k\gamma}(E)\hat{a}_{l\delta}(E)\\
%\end{align}
\begin{align}
&I_i=\frac{e}{h}\sum_{k\in \{a,b\}}\sum_{\alpha,\beta\in\{e,h\}} \sgn(\alpha)\label{paperNSN:eq:current}\\
&\times\int dE\phantom{0} \left[\delta_{ik}\delta_{\alpha\beta}-\left|s_{ik}^{\alpha\beta}\right|^2\right]f_{k\beta}(E)\notag\\
&S_{ij}=\frac{2e^2}{h}\sum_{k,l\in \{a,b\}}\sum_{\alpha,\beta,\gamma,\delta\in\{e,h\}} \sgn(\alpha)\sgn(\beta)\\
&\times\int dE\phantom{0}A_{k\gamma,l\delta}(i\alpha,E)A_{l\delta,k\gamma}(i\alpha,E)f_{k\gamma}(E)\left[1-f_{l\delta}(E)\right]\notag
\end{align}
with $ A_{k\gamma,l\delta}(i\alpha,E)=\delta_{ik}\delta_{il}\delta_{\alpha\gamma}\delta_{\alpha\delta}-s_{ik}^{\alpha\gamma\dagger}s_{il}^{\alpha\delta}$, $\sgn(\alpha=e)=1$ $\sgn(\alpha=h)=-1$, 
%$ \hat{a}^{\dagger}_{k\gamma}(E)\hat{a}_{l\delta}(E)$ are fermionic operators and there expectation
%value is $\langle\hat{a}^{\dagger}_{k\gamma}(E)\hat{a}_{l\delta}(E')\rangle=\delta_{kl}\delta_{\gamma\delta}\delta(E-E')f_{k\gamma}(E)$,
$f_{i\alpha}$ the occupancy factors for the electron and hole states in
electrode $i$, given by the Fermi function where the chemical potential are the
applied voltages 
$f_{ie}(E)=\left[1+\exp\left(\frac{E-V_i}{k_BT}\right)\right]^{-1}\xrightarrow[T\to0]{}\theta(-E+V_i)$, 
$
f_{ih}(E)=\left[1+\exp\left(\frac{E+V_i}{k_BT}\right)\right]^{-1}\xrightarrow[
T\to0]{}\theta(-E-V_i)$.

 In this one-dimensional model, both current and noise are highly sensitive to
 the distances $L_l$, $R$, $L_r$ between the barriers: They oscillate as a
 function of these distances with a period equal to the Fermi wavelength
 $\lambda_F\ll L_l, R, L_r$.  In a higher-dimensional system with more than
 one transmission mode, the oscillations in the different modes are
 independent and are thus averaged out. Multidimensional behavior can be
 simulated qualitatively with a one-dimensional system by averaging all
 quantities over one oscillation period:
%\cite{Falci,Melin-Feinberg-PRB,Melin-Martin}
\begin{align}
\label{PaperNSN:eq:av}
&\overline{X}(L_l, R, L_r)\\ &= \frac{1}{\lambda_F^3}
\int_{L_l-\frac{\lambda_F}{2}}^{L_l+\frac{\lambda_F}{2}}
dl_l\int_{R-\frac{\lambda_F}{2}}^{R+\frac{\lambda_F}{2}} dr
\int_{L_r-\frac{\lambda_F}{2}}^{L_r+\frac{\lambda_F}{2}} dr\kern2pt
X(l_l,r,l_r).\notag
\end{align}
This procedure is appropriate to describe metallic systems. These averaged
quantities are studied in section \ref{paperNNSNN:sec:AS}.

%%%%%%%%%%%%%%%%%%%%%%%%%%%%%%%%%%%%%%%%%%%%%%%%%%%%%%%%%%%%%%%%%%%%%%%%%
\section{Components of the Differential Conductance and the Differential
  Current Cross-Correlations\label{paperNNSNN:sec:comp}}
An electron, arriving from one of the normal metal reservoirs at the interface
to the superconductor, can be: i) reflected as an electron (normal reflection
(NR)), or ii) reflected as a hole (Andreev reflection (AR)), or iii)
transmitted as an electron (elastic cotunneling (EC)) or iv) transmitted as a
hole (crossed Andreev reflection (CAR)), and similarly for holes.  The
corresponding elements of the scattering matrix are for NR: $s_{aa}^{ee}$,
$s_{aa}^{hh}$, $s_{bb}^{ee}$, $s_{bb}^{hh}$, AR: $s_{aa}^{eh}$, $s_{aa}^{he}$,
$s_{bb}^{eh}$, $s_{bb}^{he}$, EC: $s_{ab}^{ee}$, $s_{ab}^{hh}$, $s_{ba}^{ee}$,
$s_{ba}^{hh}$, and CAR: $s_{ab}^{eh}$, $s_{ab}^{he}$, $s_{ba}^{eh}$,
$s_{ba}^{he}$.

The current in electrode $N_a$ given by Eq.~\eqref{paperNSN:eq:current} can
naturally be divided into AR, CAR and EC contributions (the unitarity of the
scattering matrix has been used):
\begin{widetext}
\begin{align}
 I_a=\frac{|e|}{h}\int \measure E
 &\underbrace{\left[\left(|s_{aa}^{eh}(E)|^2+|s_{aa}^{he}(E)|^2\right)(f_{ae}(E)
     - f_{ah}(E))\right.}_{\text{local Andreev
       reflection}}\notag\\ +&\underbrace{|s_{ab}^{ee}(E)|^2(f_{ae}(E) -
     f_{be}(E))+|s_{ab}^{hh}(E)|^2(f_{bh}(E)-f_{ah}(E))}_{\text{elastic
       cotunneling}}\notag\\ +&\underbrace{\left.|s_{ab}^{eh}(E)|^2(f_{ae}(E)
     -
     f_{bh}(E))+|s_{ab}^{he}(E)|^2(f_{be}(E)-f_{ah}(E))\right]}_{\text{crossed
     Andreev reflection}}.
\end{align}
In the following, we focus on i) the differential conductance in the
symmetrical case where $V_a=V_b=V$ and the current $I_a$ is differentiated
with respect to $V$, and ii) the differential nonlocal conductance in the
asymmetrical case where $V_a=0$ and the current $I_a$ is differentiated with
respect to $V_b$. In the zero temperature limit, only the nonlocal processes,
CAR and EC, contribute to the nonlocal conductance:
\begin{align}
 \left.\frac{\partial I_a}{\partial V_b}\right|_{V_a=0}
 =\underbrace{-\frac{e^2}{h}\left[
     |s_{ab}^{ee}(|e|V_b)|^2+|s_{ab}^{hh}(-|e|V_b)|^2\right]}_{\text{elastic
     cotunneling}} +\underbrace{\frac{e^2}{h}\left[|s_{ab}^{eh}(-|e|V_b)|^2 +
     |s_{ab}^{he}(|e|V_b)|^2\right]}_{\text{crossed Andreev reflection}},
\end{align}
 while the symmetric case contains local Andreev reflection and crossed
 Andreev reflection:
\begin{align}
\left.\frac{\partial I_a}{\partial V}\right|_{V_a=V_b=V} =
&\underbrace{\frac{e^2}{h}\left[\left(|s_{aa}^{eh}(|e|V)|^2 +
    |s_{aa}^{he}(|e|V)|^2\right) + \left(|s_{aa}^{eh}(-|e|V)|^2 +
    |s_{aa}^{he}(-|e|V)|^2\right)\right]}_{\text{local Andreev
    reflection}}\notag
\\ +&\underbrace{\frac{e^2}{h}\left[\left(|s_{ab}^{eh}(|e|V)|^2 +
    |s_{ab}^{he}(|e|V)|^2\right) + \left(|s_{ab}^{eh}(-|e|V)|^2 +
    |s_{ab}^{he}(-|e|V)|^2\right)\right]}_{\text{crossed Andreev reflection}}
\end{align}
Let us now perform a similar analysis for the current cross-correlations. We
study only the zero temperature limit, where
$f_{k\gamma}(E)[1-f_{l\delta}(E)]$ is zero if
$k=l$ and $\gamma=\delta$ and the current 
cross-correlations are:
\begin{equation}
 S_{ab}(T=0)=\frac{2e^2}{h}\sum_{k,l\in
   \{a,b\}}\sum_{\alpha,\beta,\gamma,\delta\in\{e,h\}}\sgn(\alpha)\sgn(\beta)\int
 dE
 s_{ak}^{\alpha\gamma\dagger}s_{al}^{\alpha\delta}s_{bl}^{\beta\delta\dagger}
 s_{bk}^{\beta\gamma}f_{k\gamma}(E)[1-f_{l\delta}(E)]
\end{equation}
\end{widetext}
Every summand in $S_{ab}$ contains the product of four elements of the
scattering matrix. As pointed out in Refs. \onlinecite{ML1992, Buettiker1990},
in difference to the situation for the current, it is impossible to combine
those matrix elements to absolute squares.  Let us now sort out and classify the
contributions of the noise as we did above for the current. We find that no
summand consists of only one kind of elements of the scattering matrix. Every
element consists of two local elements (NR or AR ) and two nonlocal elements
(CAR or EC) (see Appendix~\ref{paperNNSNN:app:Noise}). Either the two local
elements and the two nonlocal elements are identical, that gives the
components EC-NR, CAR-NR, EC-AR, CAR-AR, or all four matrix elements belong to
different categories and we will call these summands MIXED. Sometimes, it is
useful to divide MIXED further as a function of its voltage dependence.  As
the formulas for the current cross correlations are lengthy, they are
relegated into Appendix~\ref{paperNNSNN:app:Noise}.

Examination of these expressions allows an interpretation of the various
components. First, EC-NR does not involve any Andreev scattering and
corresponds to quasiparticle fluctuations across the double $NSN$
barrier. Second, CAR-NR involves two amplitudes for electron-hole scattering across $NSN$
(Crossed Andreev) and two normal scattering amplitudes. This process tracks fluctuations
of the current of split Cooper pairs emitted in- or absorbed by $S$. Third,
EC-AR involves two local Andreev scattering amplitudes and propagation of a
pair of quasiparticles in $S$. It thus reflects the fluctuations of pairs back
and forth across the $NSN$ double interface. Fourth, CAR-AR involves two
crossed Andreev and two normal Andreev amplitudes. This process
which amounts to splitting two pairs from $S$ is usually weak. Fifth, mixed processes can be
analyzed in the same fashion, they involve a combination of split
pair and quasiparticle crossing fluctuations.

For the interpretation of current cross-correlations, the global sign plays an
important role. Later on, the differential cross-correlation $\partial
S_{ab}/\partial V_{a,b}$ will be plotted. For positive applied bias voltages,
this quantity has the same sign as the current cross-correlations. For
negative applied voltages current, cross-correlations and differential current
cross-correlations have opposite signs. To avoid confusion, we only show
pictures of the differential current cross-correlations calculated for
positive bias voltages (and thus negative energies $E=-|e|V$). Due to the
electron-hole symmetry of the model, differential current cross-correlations
calculated for negative bias voltages are up to a global sign identical to the
ones calculated at positive bias voltage. For small bias voltages, current
cross-correlations depend linearly on voltage. Thus, current
cross-correlations and differential current cross-correlations show the same
qualitative behavior if studied as a function of the interface transparency or
the distance between barriers.

In Ref.~\onlinecite{FFM2010}, we performed a similar analysis of current
cross-correlations in terms of Green's functions for a NSN structure. For the
relations between these two classifications see
Appendix~\ref{PaperNSN:app:Green}.  Bignon \textit{et\,al.}~\cite{bignon} have
studied current cross-correlations in the tunneling limit. They find that
noise measurements in the tunneling limit can give access to the CAR and EC
contribution of the current.  We have just seen, that at least two processes
are involved in every component of noise, but the contributions of noise they
calculate fall into the categories EC-NR and CAR-NR. In the tunneling limit,
the NR contribution is very close to one, therefore what
remains is very similar to the current contributions. In what follows, more
generally, the CAR-NR component provides the diagnosis of Cooper pair
splitting.
%%%%%%%%%%%%%%%%%%%%%%%%%%%%%%%%%%%%%%%%%%%%%%%%%%%%%%%%%%%%%%%%%%%%%%%%%
\section{Results \label{paperNNSNN:sec:AS}}
\subsection{Positive Cross-Correlations without CAR}
If a CAR process is interpreted as the splitting of a Cooper pair into two
electrons leaving the superconductor in different electrodes, positive
cross-correlations are its logical consequence. However, the CAR process is
not the only one which can lead to positive cross-correlations. Let us
investigate in more detail the influence of the different processes on the
current cross-correlations in a NSN-system, i.\,e. in a system without the
additional barriers in the normal conducting electrodes.

\begin{figure}
 \includegraphics[width=\columnwidth]{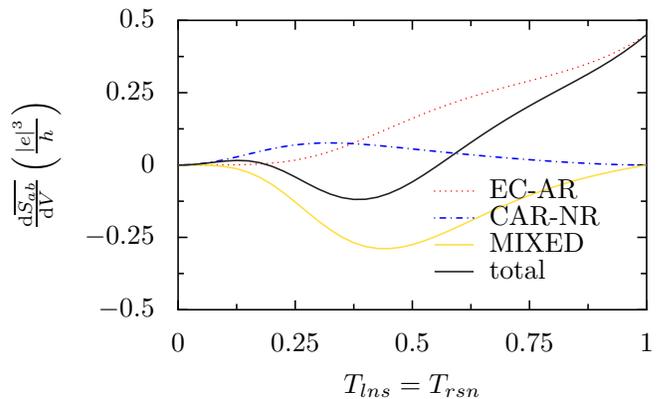}
\caption{\label{paperNSN:fig:ecnr}Averaged differential current
  cross-correlations for a symmetrical biased ($V=V_a=V_b\ll\Delta/|e|$)
  NSN-system as a function of the transparency of the interfaces
  $T_{lns}=T_{rsn}$. The positive cross-correlations at high interface
  transparency are due to the EC-AR process, represented by a dotted line.}
\end{figure}
The black line in Fig.~\ref{paperNSN:fig:ecnr} shows the averaged differential
current cross-correlations for symmetric bias ($V=V_a=V_b$). The total current
cross-correlations have already been published in Ref.~\onlinecite{FFM2010},
but here, Fig.~\ref{paperNSN:fig:ecnr} shows in addition the different parts
which contribute to the total current cross-correlations.  The total
cross-correlations are positive for high interface transparencies and for low
interface transparencies. As we have already argued in
Ref.~\onlinecite{FFM2010}, the positive cross-correlations at high interface
transparencies are not due to CAR : only processes which conserve momentum can
occur, since there are no barriers which can absorb momentum.  CAR processes
do not conserve momentum: if e.g. an electron arrives from the left-hand side
carrying momentum $k_F$, the hole that leaves at the right hand side carries
momentum $-k_F$. As the cross-correlations do not tend to zero even for very
high transparencies, they cannot be due do CAR.  Indeed, if we plot the
different components of the noise introduced in the last section separately,
we see that the positive current cross-correlations at high interface
transparencies have a different origin: a large positive EC-AR
contribution, thus correlated pair fluctuations without pair splitting. But positive current cross-correlations at low interface
transparencies are a consequence of a large CAR-NR component and therefore a
consequence of CAR processes.

We can put the contributions to the current cross-correlations into two
categories with respect to their sign, which is independent of the interface
transparency. EC-NR, CAR-AR, MIXED2 and MIXED4 carry a negative sign, CAR-NR,
EC-AR, MIXED1 and MIXED3 carry a positive sign. The current can either be
carried by electrons $I^e$ or by holes $I^h$. The sign of the different
contributions to the current cross-correlations depends on whether only
currents of the same carrier type are correlated~\cite{AD1996} ($\langle
\Delta\hat I_a^e\Delta\hat I_b^e\rangle$+$\langle \Delta\hat I_a^h\Delta\hat
I_b^h\rangle+a\leftrightarrow b$), which is the case for EC-NR, CAR-AR, MIXED2
and MIXED4 and leads to a negative sign; or whether electron currents are
correlated with hole currents ($\langle \Delta\hat I_a^e\Delta\hat
I_b^h\rangle$+$\langle \Delta\hat I_a^h\Delta\hat
I_b^e\rangle+a\leftrightarrow b$), which is the case for CAR-NR, EC-AR, MIXED1
and MIXED3 and leads to a positive sign. In purely normal conducting systems,
the electron and hole currents are uncorrelated, only correlations of the same
carrier type contribute to the current cross-correlation and lead to a
negative sign. The sign of the total current cross-correlations is a
consequence of the relative strength of the different parts of the current
cross-correlations, which depends on the interface transparency.
%%%%%%%%%%%%%%%%%%%%%%%%%%%%%%%%%%%%%%%%%%%%%%%%%%%%%%%%%%%%%%%%%%%%%%%%%
\subsection{Multiple Barriers}
In the last paragraph, we showed that positive cross-correlations due to CAR
can only be found in the tunneling regime, where the signals are quite weak.
The conductance over an NS-tunnel junction can be amplified for a ``dirty``
normal conductor containing a large number of non-magnetic impurities, where
transport is diffusive, by an effect called reflectionless tunneling
\cite{reflectionlesstunnexpt,reflectionlesstunntheo}, yielding an excess of
conductance at low energy.  It is thus natural to ask if a similar effect
could also enhance conductance and current cross-correlations in a three
terminal NSN-structure.  To answer this question within the scattering
approach, we use the model of Melsen and Beenakker \cite{MB1994} where the
disordered normal conductor is replaced by a normal conductor with an
additional tunnel barrier leading to an NNS structure.  Duhot and M\'elin
\cite{duhotmelin} have studied the influence of additional barriers on the
nonlocal conductance in three terminal NSN-structures. They indeed find that
two symmetric additional barriers enhance the nonlocal conductance. A similar
result is obtained by quasiclassical methods in Ref. \onlinecite{ZaikinPRL}.

First, let us get a deeper understanding of the result of
Ref. \onlinecite{duhotmelin} by calculating the AR, CAR and EC components of
the current separately.  Afterwards, we will study the influence of additional
barriers on the current cross-correlations.
\begin{figure}
a)\hspace{-0.5cm}\includegraphics[width=\columnwidth]{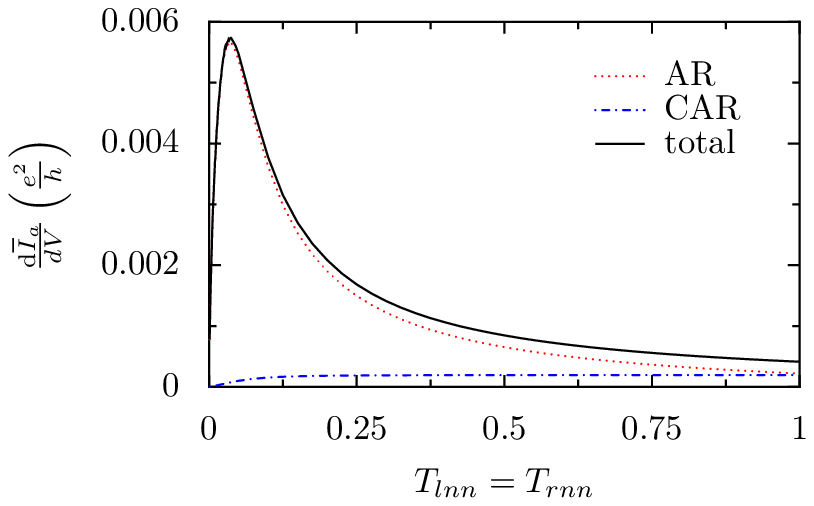}
% CondNNSNNVaVbomega1e-08.eps: 0x0 pixel, 300dpi, 0.00x0.00 cm, bb=-61 -33 231 146
b)\hspace{-0.5cm}\includegraphics[width=\columnwidth]{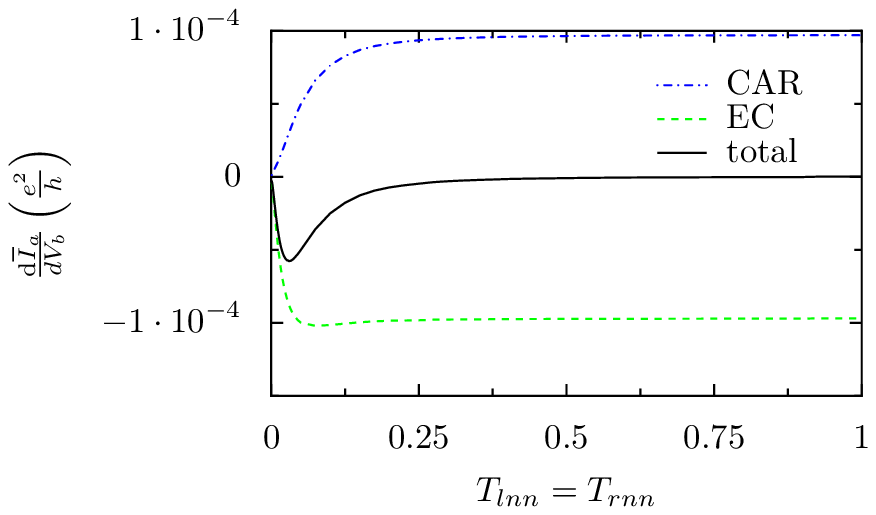}
% CondNNSNNVbAvomega1e-08.eps: 0x0 pixel, 300dpi, 0.00x0.00 cm, bb=-86 -33 231 148
\caption{\label{paperNNSNN:fig:AvCond}Averaged differential conductance in the
  limit of zero energy in a) the symmetrical bias situation
  $V_a=V_b\ll\Delta/|e|$ and b) in the asymmetrical case $V_a=0$,
  $V_b\ll\Delta/|e|$ for a superconducting electrode much shorter than the
  coherence length ($R=0.25\xi$) as a function of the transparencies of the
  additional barriers $T_{lnn}=T_{rnn}$. The barriers next to the
  superconductor are in the tunnel regime ($T_{lns}=T_{rsn}=0.01$).}
\end{figure}
\begin{figure}
a)\hspace{-0.5cm}\includegraphics[width=\columnwidth]{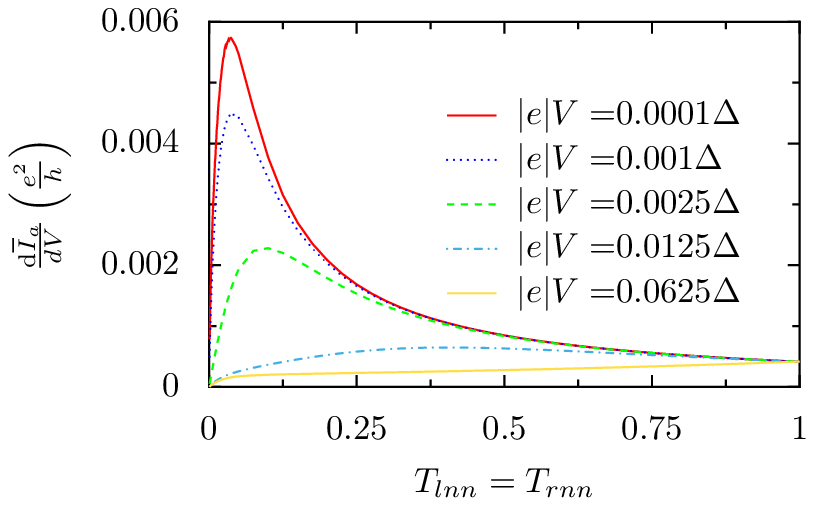}
% CondNNSNNVaVbomegaDep.eps: 0x0 pixel, 300dpi, 0.00x0.00 cm, bb=-61 -33 231 146
b)\hspace{-0.5cm}\includegraphics[width=\columnwidth]{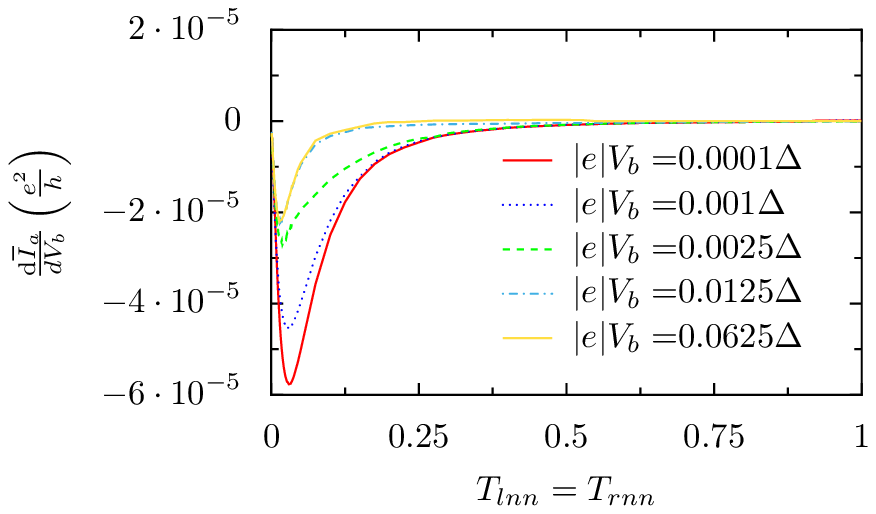}
% CondNNSNNVbAvomegaDep.eps: 0x0 pixel, 300dpi, 0.00x0.00 cm, bb=-79 -33 231 148
\caption{\label{paperNSN:fig:EnDepAvCond} Total averaged differential conductance, as in Fig.~\ref{paperNNSNN:fig:AvCond} but at different energies.}
\end{figure}
Fig.~\ref{paperNNSNN:fig:AvCond} shows the averaged conductance in the
symmetrical bias situation $V_a=V_b\ll\Delta/|e|$ and in the asymmetrical
voltage case $V_a=0$, $V_b\ll\Delta/|e|$ for a superconducting electrode much
shorter than the coherence length ($R=0.25\xi$). The sum of the AR, CAR and EC
components, traced in black, features in both cases an extremum. Yet,
examining at the behavior of these components, we see that they arise from
different mechanisms.  Let us first have a look at the symmetrically biased
case. Without the additional barriers, i.e. in the limit
$T_{lnn}=T_{rnn}\to1$, the contributions of AR and CAR are similar in
magnitude. The EC component is completely suppressed, since it is proportional
to the difference of the applied voltages. The introduction of two additional
barriers increases the AR component up to a factor 30. The shape of the curves
is similar to the one of the NNS structure derived analytically by Melsen and
Beenakker \cite{MB1994}, which can be exactly recovered by increasing the
length of the superconducting electrode far beyond the coherence
length. However, the CAR curve stays almost constant over a wide range of
values of barrier strength of the additional barriers, and it eventually
vanishes when the transparencies go to zero.

In the asymmetrical voltage case $V_a=0$, the AR component
is zero as it is proportional to the 
local voltage $V_a$. Like in the first case, the additional barriers have little influence on the CAR component, 
except for the fact that it obviously tends to zero for vanishing transparency. Over a wide range of 
barrier strength values, lowest order tunneling prevails, and the
EC component is identical in amplitude, but opposite in sign to the CAR
component \cite{falci}. For small $T_{lnn}=T_{rnn}$ values the EC
component displays a small extremum, but it is much less pronounced than the
maximum of the AR component of the 
first case. The CAR component, on the other hand, does not show any extremum. 

In the limit $T_{lnn}=T_{rnn}\to1$, the EC component tends more slowly to zero
than the CAR component which as a consequence yields a maximum in the absolute
value of the total conductance, {\it dominated by EC}. The fact that the
conductance maxima in the symmetrical bias case and in the asymmetrical bias
case have different origins can also be illustrated by studying their energy
dependence, depicted in Fig.~\ref{paperNSN:fig:EnDepAvCond}: The enhancement
of the AR component of the conductance in the symmetrical biased case
disappears completely with increasing bias voltage, as expected for a
zero-bias anomaly. On the contrary, the extremum of the conductance in the
asymmetrical biased case decreases slightly with increasing bias voltage, but
only up to a certain voltage value, then it saturates.

\begin{figure}
\includegraphics[width=\columnwidth]{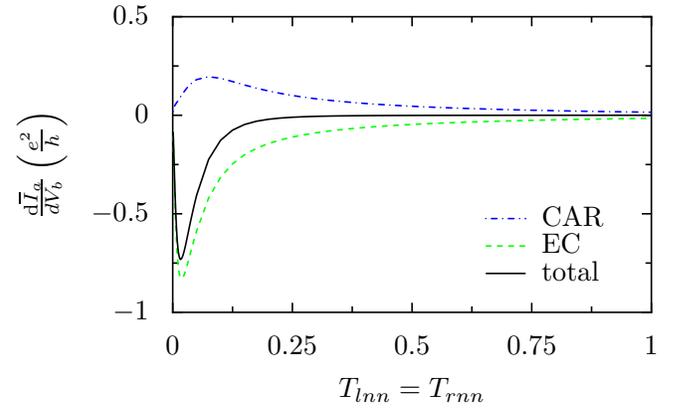}
% CondVb2Intomega1e-08.eps: 0x0 pixel, 300dpi, 0.00x0.00 cm, bb=-63 -33 231 146
\caption{\label{paperNSN:fig:Coupled}Effect of correlated averaging in the
  asymmetrical case with coupled integrals $T_{lns}=T_{rsn}=0.01$,
  $R=0.25\xi$: Now, also the EC and the CAR component are enhanced by
  reflectionless tunneling.}
\end{figure}

Why is the AR component enhanced by the additional barriers, but not the EC or
CAR components? Reflectionless tunneling occurs because the electrons and
holes resonate inside the double barrier and have therefore a higher
probability to enter the superconductor at low energy, despite
phase-averaging. In the AR case the incoming electron and the leaving hole may
encounter the same scattering path. On the contrary the EC and CAR process,
the incoming particle and the leaving particle encounter different scattering
path. The energy dependence of the conductance enhancement of the AR component
is consistent with reflectionless tunneling which occurs at low bias
voltage. At higher bias voltage, electrons and holes have different
wavevectors and the reflectionless tunneling peak disappears.  The integrals
over the phases between the additional barriers on the left- and on the
right-hand side have, of course, been taken independently. There is no reason
to think that the channel mixing, which is emulated by the integrals, on the
left- and on the right-hand side are coupled.  To verify this scenario, let us
couple the two integrals in an gedankenexperiment. We set the distance $L_l$
between the two left-hand side barriers to be equal to the distance $L_r$
between the two right-hand side barriers and perform only one integral over
$L=L_l=L_r$.  The result is shown in Fig.~\ref{paperNSN:fig:Coupled}. Now, the
CAR and the EC component are also enhanced by a large factor. Yet, the
increase of the EC component is larger than the one of the CAR component, and
EC still dominates the nonlocal conductance, like for a transparent $NSN$
structure.

To sum up the above analysis of the conductance enhancement by disorder, we
showed that the crossed processes CAR and EC cannot be amplified but by an
irrealistic correlation between disorder on the two sides of the NSN structure.
Comparing qualitatively with the different approach of Ref.
\onlinecite{ZaikinPRL}, we also find an enhancement of the crossed conductance
(Fig. 3b), which is not due to any marked maximum in CAR or EC components.
Having the sign of EC, it cannot be interpreted in terms of enhanced Cooper pair
splitting.  

\begin{figure}
a)\hspace{-0.5cm}\includegraphics[width=\columnwidth]{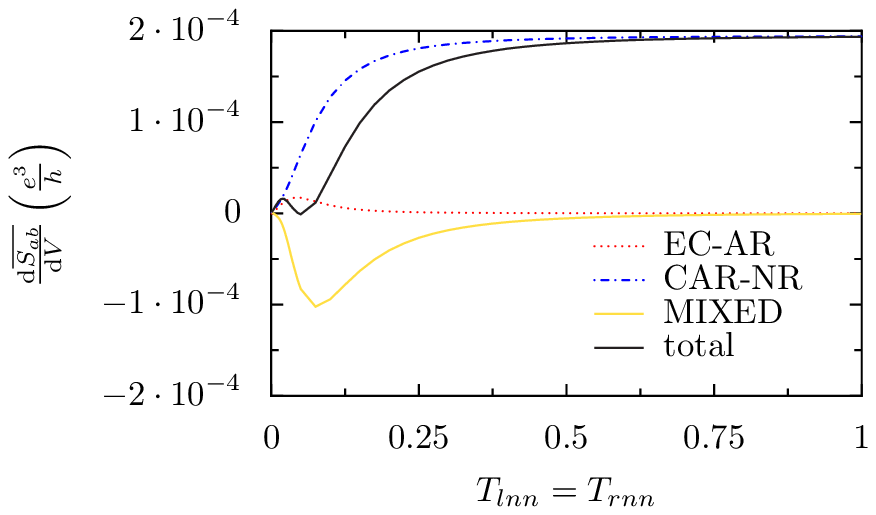}\\
b)\hspace{-0.5cm}\includegraphics[width=\columnwidth]{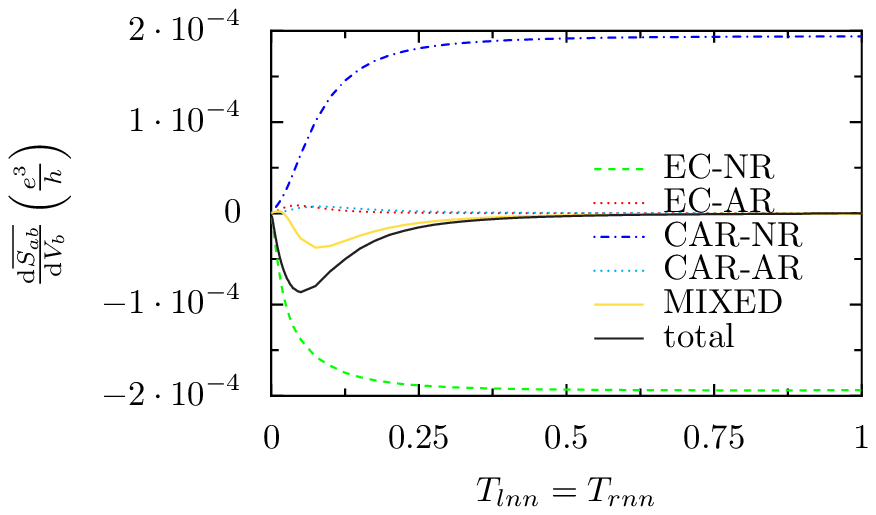}
% SabNNSNNVbAv1e-08.eps: 0x0 pixel, 300dpi, 0.00x0.00 cm, bb=-79 -33 231 148
\caption{\label{paperNSN:fig:SE-10} Averaged differential current
  cross-correlations in a) the symmetrical bias situation
  $V_a=V_b\ll\Delta/|e|$ and b) the asymmetrical bias case $V_a=0$,
  $V_b\ll\Delta/|e|$ for a superconducting electrode shorter than the
  coherence length ($R=0.25\xi$) as a function of the transparencies of the
  additional barriers $T_{lnn}=T_{rnn}$. The barriers next to the
  superconductor are in the tunnel regime ($T_{lns}=T_{rsn}=0.01$).}
\end{figure}

Let us turn back to independent averaging and consider the current
cross-correlations. The results are shown in Fig.~\ref{paperNSN:fig:SE-10}. In
the symmetrical bias case, similarly to Fig. \ref{paperNNSNN:fig:AvCond}a, the
additional barriers do not lead to an enhancement of the signal. The noise is
dominated by the CAR-NR component, featuring Cooper pair splitting and, as we
have seen above, CAR is not influenced by reflectionless tunneling.  The EC-AR
component, on the other hand, is amplified by the additional barriers, because
the AR amplitude describing a local process is amplified. This leads to a
small shoulder in the total cross-correlations. But since we are in the tunnel
regime and the leading order of CAR-NR is $T^2$ while the leading order of
EC-AR is $T^4$, the influence of the EC-AR-component is too small to lead to a
global maximum.

In the asymmetrical bias case $V_a=0,V_b\ll\Delta/|e|$, on the other hand, the
additional barriers weakly enhance the signal. But the cross-correlations are
dominated by EC-NR and are therefore negative, and they do not feature Cooper
pair splitting.  In conclusion, in a phase-averaged system, additional
barriers only enhance the AR-component, a local process. It cannot help to
amplify nonlocal signals. Again, positive cross-correlations signalling Cooper
pair splitting are only encountered in the tunneling regime. This conclusion
is contrary to the interpretation of the quasiclassical theory given in
Ref. \onlinecite{ZaikinPRBnoise}.
 
%%%%%%%%%%%%%%%%%%%%%%%%%%%%%%%%%%%%%%%%%%%%%%%%%%%%%%%%%%%%%%%%%%%%%%%%%%

\section{Conclusion}
We have found that at high transparency, crossed processes are dominated
by electron transmission and that positive cross-correlations in this range of
interface transparency are not due to Cooper pair splitting. Instead, for symmetrical voltages, 
they originate from correlated fluctuations of Cooper pairs from the superconductor 
to both metallic contacts and vice-versa. Cooper pair
splitting in the tunnel regime cannot be enhanced with additional barriers by
a process similar to reflectionless tunneling, if an average (here, over the
interbarrier lengths) has to be performed, mimicking disorder landscapes which
are uncorrelated on the two sides of the set-up. In analogy, one expects that
the same conclusion holds if one uses diffusive normal metals. These
conclusions are important for settling future experimental programs. Positive
cross-correlations might well be observed at high transparency, but they are
not a signature of Cooper pair splitting. They are not related to spatially
separated spin-entangled pairs. 
%%%%%%%%%%%%%%%%%%%%%%%%%%%%%%%%%%%%
%%%%%%%%%%%%% CHANGE 2 %%%%%%%%%%%%%
%%%%%%%%%%%%%%%%%%%%%%%%%%%%%%%%%%%%
Conductance and cross-correlation measurements with controlled and tunable interface transparencies would be very useful, and 
may be attempted for instance in carbon nanotubes junctions.
Finally, a cross-over to negative current cross-correlations for a highly
transparent NSN junction biased above the gap is expected. Its theoretical
description is more involved, because it requires taking nonequilibrium
effects such as charge imbalance \cite{imb1,imb2,imb3} into account. A
  starting point for those calculations can be found in
  Ref.~\onlinecite{Melin-Bergeret-Yeyati}.
%%%%%%%%%%%%%%%%%%%%%%%%%%%%%%%%%%%%%%%%%%%
%%%%%%%%%%%%% END OF CHANGE 2 %%%%%%%%%%%%%
%%%%%%%%%%%%%%%%%%%%%%%%%%%%%%%%%%%%%%%%%%%

\section*{Acknowledgements}
The authors have benefited from several fruitful discussions with
B. Dou\c{c}ot. Part of this work was supported by ANR Project "Elec-EPR".

%%%%%%%%%%%%%%%%%%%%%%%%%%%%%%%%%%%%%%%%%%%%%%%%%%%%%%%%%%%%%%%%%%%%%%%%%%%
\appendix 
\section{Details on the scattering approach\label{paperNSN:app:BTK}}
The elements $s_{ij}^{\alpha\beta}$ of the scattering matrix are calculated
within the BTK approach\cite{BTK1982}: Two-component wavefunctions, where the
upper component describes electrons and the lower components holes, are
matched at the interfaces and the coefficients of the resulting system of
equations give the elements of the scattering matrix.

In the normal conductors the the wavefunctions are plane waves with
wavevectors close to the Fermi wavevector $\hbar k_F=\sqrt{2m\mu}$. The
wavevector for electrons reads $\hbar q^{+}=\sqrt{2m}\sqrt{\mu+E}$, the one
for holes $\hbar q^{+}=\sqrt{2m}\sqrt{\mu-E}$. In the superconductor, the
wavefunction has to obey the Bogoliubov-De Gennes equation, where the
superconducting gap is supposed to be a positive constant inside the
superconductor and zero outside of it. This is achieved by modifying the
amplitudes of the wavefunction in the superconductor with the coherence
factors $u_{E}$ and $v_{E}$ which read for energies smaller than the gap
($E<\Delta$):
\begin{align}
 u_{E}=\frac{1}{\sqrt{2}}\sqrt{1+\frac{i\sqrt{\Delta^2-E^2}}{E}},\phantom{000} 
v_{E}=\frac{1}{\sqrt{2}}\sqrt{1-\frac{i\sqrt{\Delta^2-E^2}}{E}}
\end{align}
and by using for quasi-particles proportional to $\bp u_E\\v_E\ep$ the
wavevector $\hbar k^{+}=\sqrt{2m}\sqrt{\mu+i\sqrt{\Delta^2-E^2}}$ and for
quasi-particles proportional to $\bp v_E\\u_E\ep$ the wavevector $\hbar
k^{-}=\sqrt{2m}\sqrt{\mu-i\sqrt{\Delta^2-E^2}}$.

For example, the wavefunctions for an electron incoming from electrode N$_a$
take the form
\begin{widetext}
\begin{align}
\displaybreak[0]
\psi_{\mathrm{N}_a}(x)&=
\begin{pmatrix}1\\0\end{pmatrix}
\left(1\phantom{0}e^{iq^{+}x}+s^{ee}_{aa}\phantom{0} e^{-iq^{+}x}\right)
+ \begin{pmatrix}0\\1\end{pmatrix} 
\left(s^{he}_{aa}\phantom{0}e^{iq^{-}x}+0\phantom{0}e^{-iq^{-}x}\right),\\
\displaybreak[0]
\psi_{\mathrm{N}_l}(x)&=
\begin{pmatrix}1\\0\end{pmatrix}\left(c_1\phantom{0}e^{iq^{+}x}
+c_2\phantom{0}e^{-iq^{+}(x-L_l)}\right)+\begin{pmatrix}0\\1\end{pmatrix}\left(c_3\phantom{0} e^{iq^{-}x}
+c_4\phantom{0} e^{-iq^{-}(x-L_l)}\right),\\
\displaybreak[0]
\psi_{\mathrm{S}}(x)&=
\begin{pmatrix}u_E\\v_E \end{pmatrix} \left( c_5\phantom{0} e^{ik^{+}(x-L_l)}
+c_6\phantom{0} e^{-ik^{+}(x-L_l-R)}\right)\notag\\
&+\begin{pmatrix}v_E\\u_E \end{pmatrix} \left( c_7\phantom{0} e^{-ik^{-}(x-L_l}
+c_8\phantom{0} e^{ik^{-}(x-L_l-R)}\right),\\
\psi_{\mathrm{N}_r}(x)&=
\begin{pmatrix}1\\0\end{pmatrix}\left(c_9\phantom{0}e^{iq^{+}(x-L_l-R)}
+c_{10}\phantom{0}e^{-iq^{+}(x-L_l-R-L_r)}\right)\notag\\
&+\begin{pmatrix}0\\1\end{pmatrix}\left(c_{11}\phantom{0} e^{iq^{-}(x-L_l-R)}
+c_{12}\phantom{0} e^{-iq^{-}(x-L_l-R-L_r)}\right),\\
\displaybreak[0]
\psi_{\mathrm{N}_b}(x)&=
\begin{pmatrix}1\\0\end{pmatrix}\left(s^{e,e}_{b,a}\phantom{0}e^{iq^{+}(x-L_l-R-L_r)}
+0\phantom{0}e^{-iq^{+}(x-L_l-R-L_r)}\right)\notag\\
&+\begin{pmatrix}0\\1\end{pmatrix}\left(0\phantom{0} e^{iq^{-}(x-L_l-R)}
+s^{h,e}_{b,a}\phantom{0} e^{-iq^{-}(x-L_l-R-L_r)}\right)
\end{align}n
\end{widetext}
in the sections N$_a$, N$_l$, S, N$_r$, and N$_b$ respectively [see
  Fig.~\ref{paperNNSNN:fig:model}] and give access to the scattering matrix
elements $s^{e,e}_{a,a}$, $s^{h,e}_{a,a}$, $s^{h,e}_{b,a}$ and
$s^{e,e}_{b,a}$. The remaining elements of the scattering matrix can be
obtained from the other possible scattering processes i.e. a hole incoming
from electrode N$_a$, an electron/hole incoming from electrode N$_b$.

The interfaces are modeled by $\delta$-potentials $V(x)=Z\hbar v_F\delta(x)$,
where the BTK parameter $Z$ is connected to the interface transparency $T$ by
$T=(1+Z^2)^{-1}$.
%$Z=H/\hbar v_F$ .
The elements of the scattering matrix can be determined and the constants
$c_i$ eliminated using the continuity of the wavefunctions at the interfaces
[$\psi_{N_a}(0)=\psi_{N_l}(0)$ etc.] and the boundary condition for the
derivatives [$\psi_{N_l}'(0)-\psi_{N_a}'(0) = Z\hbar v_F \psi_a(0)$ etc.] at
every interface.

In simple cases, i. e. for only two or three sections and in the limit of zero
energy, the system of equations giving the scattering matrix elements can be
solved analytically (see \cite{FFM2010}), but in the present case of five
sections the expressions become so unhandy that the equations are solved
numerically.
\section{Components of Current Cross Correlations\label{paperNNSNN:app:Noise}}
\begin{widetext}
Components of the current cross-correlations:
\tiny
\begin{align*}
&S_{ab}(T=0,V_a,V_b)=\frac{2e^2}{h}\int dE \phantom{0}\biggr(&&\\
&\left.
\begin{aligned}
&  2\Re\left[s_{ab}^{ee} s_{ba}^{ee} s_{aa}^{ee\dagger} s_{bb}^{ee\dagger}\right](\theta(|e|V_a-E)-2\theta(|e|V_a-E)\theta(|e|V_b-E)+\theta(|e|V_b-E))\\
+& 2\Re \left[s_{ab}^{hh} s_{ba}^{hh} s_{aa}^{hh\dagger} s_{bb}^{hh\dagger} \right](\theta(-|e|V_a-E)-2\theta(-|e|V_a-E)\theta(-|e|V_b-E)+\theta(-|e|V_b-E))
\end{aligned}\right\}{EC-NR}&&\displaybreak[0]\\ 
&\left.
\begin{aligned}
+& 2\Re \left[ s_{ba}^{eh} s_{ab}^{he} s_{aa}^{hh\dagger} s_{bb}^{ee\dagger}\right] (-\theta(-|e|V_a-E)+2\theta(-|e|V_a-E)\theta(|e|V_b-E)-\theta(|e|V_b-E))\\
+& 2\Re \left[ s_{ab}^{eh} s_{ba}^{he} s_{aa}^{ee\dagger} s_{bb}^{hh\dagger}\right] (-\theta(|e|V_a-E)+2\theta(|e|V_a-E)\theta(-|e|V_b-E)-\theta(-|e|V_b-E))\\
\end{aligned}\right\}{CAR-NR}&&\displaybreak[0]\\
&\left.
\begin{aligned}
+&  2\Re \left[s_{ab}^{hh} s_{ba}^{ee} s_{bb}^{eh\dagger} s_{aa}^{he\dagger} \right] (-\theta(|e|V_a-E)+2\theta(|e|V_a-E)\theta(-|e|V_b-E)-\theta(-|e|V_b-E))\\
+&  2\Re \left[s_{ab}^{ee} s_{ba}^{hh} s_{aa}^{eh\dagger} s_{bb}^{he\dagger} \right] (-\theta(-|e|V_a-E)+2\theta(-|e|V_a-E)\theta(|e|V_b-E)-\theta(|e|V_b-E))\\
\end{aligned}\right\}{EC-AR}&&\displaybreak[0]\\
&\left.
\begin{aligned}
+&  2\Re \left[s_{ba}^{he} s_{ab}^{he} s_{aa}^{he\dagger} s_{bb}^{he\dagger}\right] (\theta(|e|V_a-E)-2\theta(|e|V_a-E)\theta(|e|V_b-E)+\theta(|e|V_b-E))\\
+&  2\Re \left[s_{ab}^{eh} s_{ba}^{eh} s_{aa}^{eh\dagger} s_{bb}^{eh\dagger}\right] (\theta(-|e|V_a-E)-2\theta(-|e|V_a-E)\theta(-|e|V_b-E)+\theta(-|e|V_b-E))\\
\end{aligned}\right\}{CAR-AR}&&\displaybreak[0]\\ 
&\left.
\begin{aligned}
+&  2\Re \left[s_{ab}^{eh} s_{ba}^{hh} s_{bb}^{hh\dagger} s_{aa}^{eh\dagger}+ s_{ab}^{hh} s_{ba}^{eh} s_{aa}^{hh\dagger} s_{bb}^{eh\dagger}\right] (-\theta(-|e|V_a-E)+2\theta(-|e|V_a-E)\theta(-|e|V_b-E)-\theta(-|e|V_b-E))\\
+&  2\Re \left[s_{ab}^{ee} s_{ba}^{he} s_{aa}^{ee\dagger} s_{bb}^{he\dagger}+ s_{ba}^{ee} s_{ab}^{he} s_{bb}^{ee\dagger} s_{aa}^{he\dagger}\right] (-\theta(|e|V_a-E)+2\theta(|e|V_a-E)\theta(|e|V_b-E)-\theta(|e|V_b-E))\\
\end{aligned}\right\}{MIXED1} &&\displaybreak[0]\\ 
&\left.
\begin{aligned}
+&  2\Re \left[s_{ab}^{ee} s_{ba}^{eh} s_{bb}^{ee\dagger} s_{aa}^{eh\dagger}+s_{ba}^{hh} s_{ab}^{he} s_{aa}^{hh\dagger} s_{bb}^{he\dagger} \right] (\theta(-|e|V_a-E)-2\theta(-|e|V_a-E)\theta(|e|V_b-E)+\theta(|e|V_b-E))\\
+&  2\Re \left[s_{ab}^{eh} s_{ba}^{ee} s_{aa}^{ee\dagger} s_{bb}^{eh\dagger}+s_{ab}^{hh} s_{ba}^{he} s_{bb}^{hh\dagger} s_{aa}^{he\dagger} \right] (\theta(|e|V_a-E)-2\theta(|e|V_a-E)\theta(-|e|V_b-E)+\theta(-|e|V_b-E))\\
\end{aligned}\right\}{MIXED2} &&\displaybreak[0]\\ 
&\left.
\begin{aligned}
+&  2\Re \left[s_{aa}^{hh} s_{ba}^{ee} s_{ba}^{eh\dagger} s_{aa}^{he\dagger}+s_{aa}^{ee} s_{ba}^{hh} s_{ba}^{he\dagger} s_{aa}^{eh\dagger} \right] (-\theta(|e|V_a-E)+2\theta(|e|V_a-E)\theta(-|e|V_a-E)-\theta(-|e|V_a-E))\\
\end{aligned}\right\}{MIXED3a} &&\displaybreak[0]\\ 
&\left.
\begin{aligned}
+&  2\Re \left[s_{ab}^{eh} s_{bb}^{he} s_{ab}^{ee\dagger} s_{bb}^{hh\dagger}+s_{ab}^{hh} s_{bb}^{ee} s_{ab}^{he\dagger} s_{bb}^{eh\dagger}\right] (-\theta(|e|V_b-E)+2\theta(|e|V_b-E)\theta(-|e|V_b-E)-\theta(-|e|V_b-E))\\ 
\end{aligned}\right\}{MIXED3b} &&\displaybreak[0]\\ 
&\left.
\begin{aligned}
+&  2\Re \left[s_{aa}^{eh} s_{ba}^{ee} s_{aa}^{ee\dagger} s_{ba}^{eh\dagger}+s_{ba}^{hh} s_{aa}^{he} s_{aa}^{hh\dagger}s_{ba}^{he\dagger} \right] (\theta(|e|V_a-E)-2\theta(|e|V_a-E)\theta(-|e|V_a-E)+\theta(-|e|V_a-E))\\
\end{aligned}\right\}{MIXED4a} &&\displaybreak[0]\\ 
&\left.
\begin{aligned}
+&  2\Re \left[s_{ab}^{ee} s_{ab}^{eh} s_{bb}^{ee\dagger} s_{ab}^{eh\dagger})+s_{ab}^{eh} s_{bb}^{he} s_{bb}^{hh\dagger} s_{ab}^{he\dagger})\right] (\theta(|e|V_b-E)-2\theta(|e|V_b-E)\theta(-|e|V_b-E)+\theta(-|e|V_b-E))\biggl)\\
\end{aligned}\right\}{MIXED4b} &&
\end{align*}
\normalsize
Differential current cross-correlations in the nonlocal conductance setup:
\tiny
\begin{align*}
&\frac{\measure S_{ab}(T=0,V_a=0,V_b)}{\measure V_b}=\left.\right.\frac{2|e|^3}{h}\sgn(V_b)\biggl(\\
&\left.
\begin{aligned}
&\phantom{+}  2\Re\left[s_{ab}^{ee}(|e|V_b) s_{ba}^{ee}(|e|V_b) s_{aa}^{ee\dagger}(|e|V_b) s_{bb}^{ee\dagger}(|e|V_b)\right]
+ 2\Re \left[s_{ab}^{hh}(-|e|V_b) s_{ba}^{hh}(-|e|V_b) s_{aa}^{hh\dagger}(-|e|V_b) s_{bb}^{hh\dagger}(-|e|V_b)\right]\\
\end{aligned}\right\}{EC-NR}\displaybreak[0]\\ 
&\left.
\begin{aligned}
-& 2\Re \left[ s_{ba}^{eh}(|e|V_b) s_{ab}^{he}(|e|V_b) s_{aa}^{hh\dagger}(|e|V_b) s_{bb}^{ee\dagger}(|e|V_b)\right]
-  2\Re \left[ s_{ab}^{eh}(-|e|V_b) s_{ba}^{he}(-|e|V_b) s_{aa}^{ee\dagger}(-|e|V_b) s_{bb}^{hh\dagger}(-|e|V_b)\right]\\
\end{aligned}\right\}{CAR-NR}\displaybreak[0]\\
&\left.
\begin{aligned}
-&  2\Re \left[s_{ab}^{ee}(|e|V_b) s_{ba}^{hh}(|e|V_b) s_{aa}^{eh\dagger}(|e|V_b) s_{bb}^{he\dagger} (|e|V_b)\right]
-  2\Re \left[s_{ab}^{hh}(-|e|V_b) s_{ba}^{ee}(-|e|V_b) s_{bb}^{eh\dagger}(-|e|V_b) s_{aa}^{he\dagger}(-|e|V_b) \right]\\
\end{aligned}\right\}{EC-AR}\displaybreak[0]\\
&\left.
\begin{aligned}
+&  2\Re \left[s_{ba}^{he}(|e|V_b) s_{ab}^{he}(|e|V_b) s_{aa}^{he\dagger}(|e|V_b) s_{bb}^{he\dagger}(|e|V_b)\right]
+  2\Re \left[s_{ab}^{eh} (-|e|V_b)s_{ba}^{eh} (-|e|V_b)s_{aa}^{eh\dagger}(-|e|V_b) s_{bb}^{eh\dagger}(-|e|V_b)\right]\\
\end{aligned}\right\}{CAR-AR}\displaybreak[0]\\ 
&\left.
\begin{aligned}
-&  2\Re \left[s_{ab}^{ee}(|e|V_b) s_{ba}^{he}(|e|V_b) s_{aa}^{ee\dagger}(|e|V_b) s_{bb}^{he\dagger}(|e|V_b)+ s_{ba}^{ee}(|e|V_b) s_{ab}^{he}(|e|V_b) s_{bb}^{ee\dagger}(|e|V_b) s_{aa}^{he\dagger}(|e|V_b)\right]\\
-&  2\Re \left[s_{ab}^{eh}(-|e|V_b) s_{ba}^{hh}(-|e|V_b) s_{bb}^{hh\dagger}(-|e|V_b) s_{aa}^{eh\dagger}(-|e|V_b)+ s_{ab}^{hh}(-|e|V_b) s_{ba}^{eh}(-|e|V_b) s_{aa}^{hh\dagger}(-|e|V_b) s_{bb}^{eh\dagger}(-|e|V_b)\right]\\
\end{aligned}\right\}{MIXED1} &&\displaybreak[0]\\ 
&\left.
\begin{aligned}
+&  2\Re \left[s_{ab}^{eh}(|e|V_b) s_{ba}^{ee}(|e|V_b) s_{aa}^{ee\dagger}(|e|V_b) s_{bb}^{eh\dagger}(|e|V_b)+s_{ab}^{hh}(|e|V_b) s_{ba}^{he}(|e|V_b) s_{bb}^{hh\dagger}(|e|V_b) s_{aa}^{he\dagger}(|e|V_b) \right]\\
+&  2\Re \left[s_{ab}^{ee}(-|e|V_b) s_{ba}^{eh}(-|e|V_b) s_{bb}^{ee\dagger}(-|e|V_b) s_{aa}^{eh\dagger}(-|e|V_b)+s_{ba}^{hh}(-|e|V_b) s_{ab}^{he}(-|e|V_b) s_{aa}^{hh\dagger}(-|e|V_b) s_{bb}^{he\dagger}(-|e|V_b) \right]\\
\end{aligned}\right\}{MIXED2} &&\displaybreak[0]\\ 
&\left.
\begin{aligned}
-&  2\Re \left[s_{ab}^{eh}(|e|V_b) s_{bb}^{he}(|e|V_b) s_{ab}^{ee\dagger}(|e|V_b) s_{bb}^{hh\dagger}(|e|V_b)+s_{ab}^{hh}(|e|V_b) s_{bb}^{ee}(|e|V_b) s_{ab}^{he\dagger}(|e|V_b) s_{bb}^{eh\dagger}(|e|V_b)\right]\\
-&  2\Re \left[s_{ab}^{eh}(-|e|V_b) s_{bb}^{he}(-|e|V_b) s_{ab}^{ee\dagger}(-|e|V_b) s_{bb}^{hh\dagger}(-|e|V_b)+s_{ab}^{hh}(-|e|V_b) s_{bb}^{ee}(-|e|V_b) s_{ab}^{he\dagger}(-|e|V_b) s_{bb}^{eh\dagger}(-|e|V_b)\right]\\
\end{aligned}\right\}{MIXED3b} &&\displaybreak[0]\\ 
&\left.
\begin{aligned}
+&  2\Re \left[s_{ab}^{ee}(|e|V_b) s_{ab}^{eh}(|e|V_b) s_{bb}^{ee\dagger}(|e|V_b) s_{ab}^{eh\dagger}(|e|V_b)+s_{ab}^{eh}(|e|V_b) s_{bb}^{he}(|e|V_b) s_{bb}^{hh\dagger}(|e|V_b) s_{ab}^{he\dagger}(|e|V_b)\right]\\
+&  2\Re \left[s_{ab}^{ee}(-|e|V_b) s_{ab}^{eh}(-|e|V_b) s_{bb}^{ee\dagger}(-|e|V_b) s_{ab}^{eh\dagger}(-|e|V_b)+s_{ab}^{eh}(-|e|V_b) s_{bb}^{he}(-|e|V_b) s_{bb}^{hh\dagger}(-|e|V_b) s_{ab}^{he\dagger}(-|e|V_b)\right]\biggl)\\
\end{aligned}\right\}{MIXED4b} &&
\end{align*}
\normalsize
Differential current cross-correlations in the symmetrical setup:
\tiny
\begin{align*}
&\frac{dS_{ab}(T=0,V_a+V,V_b=V)}{dV}=\frac{2|e|^3}{h}\sgn(|e|V)\phantom{0}\biggr(\\
&\left.
\begin{aligned}
-& 2\Re \left[ s_{ba}^{eh}(|e|V) s_{ab}^{he}(|e|V) s_{aa}^{hh\dagger}(|e|V) s_{bb}^{ee\dagger}(|e|V)+s_{ab}^{eh}(|e|V) s_{ba}^{he}(|e|V) s_{aa}^{ee\dagger}(|e|V) s_{bb}^{hh\dagger}(|e|V)\right]\\
-& 2\Re \left[ s_{ba}^{eh}(-|e|V) s_{ab}^{he}(-|e|V) s_{aa}^{hh\dagger}(-|e|V) s_{bb}^{ee\dagger}(-|e|V)+s_{ab}^{eh}(-|e|V) s_{ba}^{he}(-|e|V) s_{aa}^{ee\dagger}(-|e|V) s_{hh}^{bb\dagger}(-|e|V)\right]\\
\end{aligned}\right\}{CAR-NR}\displaybreak[0]\\
&\left.
\begin{aligned}
-&  2\Re \left[s_{ab}^{hh}(|e|V) s_{ba}^{ee}(|e|V) s_{bb}^{eh\dagger}(|e|V) s_{aa}^{he\dagger}(|e|V)+s_{ab}^{ee}(|e|V) s_{ba}^{hh}(|e|V) s_{aa}^{eh\dagger}(|e|V) s_{bb}^{he\dagger}(|e|V) \right]\\
-&  2\Re \left[s_{ab}^{hh}(-|e|V) s_{ba}^{ee}(-|e|V) s_{bb}^{eh\dagger}(-|e|V) s_{aa}^{he\dagger}(-|e|V)+s_{ab}^{ee}(-|e|V) s_{ba}^{hh}(-|e|V) s_{aa}^{eh\dagger}(-|e|V) s_{bb}^{he\dagger}(-|e|V) \right]\\
\end{aligned}\right\}{EC-AR}\displaybreak[0]\\
&\left.
\begin{aligned}
+&  2\Re \left[s_{ab}^{ee}(|e|V) s_{ba}^{eh}(|e|V) s_{bb}^{ee\dagger}(|e|V) s_{aa}^{eh\dagger}(|e|V)+s_{ba}^{hh}(|e|V) s_{ab}^{he}(|e|V) s_{aa}^{hh\dagger}(|e|V) s_{bb}^{he\dagger}(|e|V) \right]\\
+&  2\Re \left[s_{ab}^{ee}(-|e|V) s_{ba}^{eh}(-|e|V) s_{bb}^{ee\dagger}(-|e|V) s_{aa}^{eh\dagger}(-|e|V)+s_{ba}^{hh}(-|e|V) s_{ab}^{he}(-|e|V) s_{aa}^{hh\dagger}(-|e|V) s_{bb}^{he\dagger}(-|e|V) \right]\\
+&  2\Re \left[s_{ab}^{eh}(|e|V) s_{ba}^{ee}(|e|V) s_{aa}^{ee\dagger}(|e|V) s_{bb}^{eh\dagger}(|e|V)+s_{ab}^{hh}(|e|V) s_{ba}^{he}(|e|V) s_{bb}^{hh\dagger}(|e|V) s_{aa}^{he\dagger}(|e|V) \right]\\
+&  2\Re \left[s_{ab}^{eh}(-|e|V) s_{ba}^{ee}(-|e|V) s_{aa}^{ee\dagger}(-|e|V) s_{bb}^{eh\dagger}(-|e|V)+s_{ab}^{hh}(-|e|V) s_{ba}^{he}(-|e|V) s_{bb}^{hh\dagger}(-|e|V) s_{aa}^{he\dagger}(-|e|V) \right]\\
\end{aligned}\right\}{MIXED2} &&\displaybreak[0]\\ 
&\left.
\begin{aligned}
-&  2\Re \left[s_{aa}^{hh}(|e|V) s_{ba}^{ee}(|e|V) s_{ba}^{eh\dagger}(|e|V) s_{aa}^{he\dagger}(|e|V)+s_{aa}^{ee}(|e|V) s_{ba}^{hh}(|e|V) s_{ba}^{he\dagger}(|e|V) s_{aa}^{eh\dagger}(|e|V) \right]\\
-&  2\Re \left[s_{aa}^{hh}(-|e|V) s_{ba}^{ee}(-|e|V) s_{ba}^{eh\dagger}(-|e|V) s_{aa}^{he\dagger}(-|e|V)+s_{aa}^{ee}(-|e|V) s_{ba}^{hh}(-|e|V) s_{ba}^{he\dagger}(-|e|V) s_{aa}^{eh\dagger}(-|e|V)\right]\\
\end{aligned}\right\}{MIXED3a} &&\displaybreak[0]\\ 
&\left.
\begin{aligned}
-&  2\Re \left[s_{ab}^{eh}(|e|V) s_{bb}^{he}(|e|V) s_{ab}^{ee\dagger}(|e|V) s_{bb}^{hh\dagger}(|e|V)+s_{ab}^{hh}(|e|V) s_{bb}^{ee}(|e|V) s_{ab}^{he\dagger}(|e|V) s_{bb}^{eh\dagger}(|e|V)\right]\\
-&  2\Re \left[s_{ab}^{eh}(-|e|V) s_{bb}^{he}(-|e|V) s_{ab}^{ee\dagger}(-|e|V) s_{bb}^{hh\dagger}(-|e|V)+s_{ab}^{hh}(-|e|V) s_{bb}^{ee}(-|e|V) s_{ab}^{he\dagger}(-|e|V) s_{bb}^{eh\dagger}(-|e|V)\right]\\
\end{aligned}\right\}{MIXED3b} &&\displaybreak[0]\\ 
&\left.
\begin{aligned}
+&  2\Re \left[s_{aa}^{eh}(|e|V) s_{ba}^{ee}(|e|V) s_{aa}^{ee\dagger}(|e|V) s_{ba}^{eh\dagger}(|e|V)+s_{ba}^{hh}(|e|V) s_{aa}^{he}(|e|V) s_{aa}^{hh\dagger}(|e|V)s_{ba}^{he\dagger}(|e|V) \right]\\
+&  2\Re \left[s_{aa}^{eh}(-|e|V) s_{ba}^{ee}(-|e|V) s_{aa}^{ee\dagger}(-|e|V) s_{ba}^{eh\dagger}(-|e|V)+s_{ba}^{hh}(-|e|V) s_{aa}^{he}(-|e|V) s_{aa}^{hh\dagger}(-|e|V)s_{ba}^{he\dagger}(-|e|V)\right]\\
\end{aligned}\right\}{MIXED4a} &&\displaybreak[0]\\ 
&\left.
\begin{aligned}
+&  2\Re \left[s_{ab}^{ee}(|e|V) s_{ab}^{eh}(|e|V) s_{bb}^{ee\dagger}(|e|V) s_{ab}^{eh\dagger}(|e|V)+s_{ab}^{eh}(|e|V) s_{bb}^{he}(|e|V) s_{bb}^{hh\dagger}(|e|V) s_{ab}^{he\dagger}(|e|V)\right]\\
+&  2\Re \left[s_{ab}^{ee}(-|e|V) s_{ab}^{eh}(-|e|V) s_{bb}^{ee\dagger}(-|e|V) s_{ab}^{eh\dagger}(-|e|V)+s_{ab}^{eh}(-|e|V) s_{bb}^{he}(-|e|V) s_{bb}^{hh\dagger}(-|e|V) s_{ab}^{he\dagger}(-|e|V)\right]\biggl)\\
\end{aligned}\right\}{MIXED4b}
\end{align*}
\normalsize
\end{widetext}
%%%%%%%%%%%%%%%%%%%%%%%%%%%%%%%%%%%%%%%%%%%%%%%%%%%%%%%%%%%%%%%%%%%%%%%%%%%%%%%%%%%%%%%%%%%%%%%%%%%%%%%%%%%%%%%%%%%%%%%%%%%%%%%%%%%%%%%%%%%%%%%%%%%%%%%%%%%%%%%%%%%%%%%%%%%%%%%%%%%%%%%%%%%%%%%%%%%
\section{Relations between the noise classification in the BTK and in the
Green's functions approach}
\label{PaperNSN:app:Green}

\begin{table}
\begin{tabular}{l|l}
Scattering matrix& Green's function\\
classification& classification\\
\hline
\hline
CAR-NR & CAR\\
\hline
EC-AR & \ensuremath {\mathrm {AR}\text{-}\overline{\mathrm{AR}}}\\
\hline
MIXED1, MIXED2 & PRIME\\
\hline
EC-NR &EC\\
\hline
CAR-AR & AR-AR\\
\hline
MIXED3, MIXED4 & MIXED
\end{tabular}
\caption{\label{PaperNSN:tab:correspondence} Correspondences between the
categories in the language of Green's functions from~\cite{FFM2010} and in the language of
scattering matrix elements.}
\end{table}

The elements of the scattering-matrix are connected to the
retarded Green's functions of the tight binding model studied in \cite{FFM2010}
via
\begin{align}
 s_{ij}^{\alpha\beta}=i\delta_{ij}+2\pi t_i t_j
\sqrt{\rho_i^{\alpha}}\sqrt{\rho_j^{\beta}}G_{ij\alpha\beta}^R
\end{align}
where $t_i$ is the transmission coefficient of the barrier $i$,
$\rho_i^{\alpha}$ the density of electron or hole states of electrode $i$ and
$G_{ij\alpha\beta}^R$ the Green's function connecting the first site in the
superconductor next to the electrode $j$ to the first site in the
superconductor next to the electrode
$i$.

Table~\ref{PaperNSN:tab:correspondence} shows the correspondences between the
categories in the language of Green's functions and in the language of
scattering matrix elements.
%

%%%%%%%%%%%%%%%%%%%%%%%%%%%%%%%%%%%%%%%%%%%%%%%%%%%%%%%%%%%%%%%%%%%%%%%%%%%
%\bibliographystyle{unsrt}
%\bibliography{biblioNSN}

\end{document}